\documentclass[twoside,english]{elsart}

\usepackage{amssymb}
\usepackage{graphicx}
\usepackage{babel}
\usepackage{multicol}
\usepackage[centerlast]{caption2}
\usepackage{setspace}

\newcommand{\registered}{\circledR}

\begin{document}

\begin{frontmatter}

\title{Quantum efficiency of technical metal photocathodes
under laser irradiation of various wavelength}

\author[PSI]{F. Le Pimpec\corauthref{cor1}},
\ead{frederic.le.pimpec@psi.ch}
\author[PSI,EPFL]{F. Ardana-Lamas},
\author[PSI,EPFL]{C.P. Hauri},
\author[PSI,EPFL]{C. Milne}
\address[PSI]{Paul Scherrer Institute \\5232 Villigen
Switzerland}
\address[EPFL]{Ecole Polytechnique F\'ed\'erale de Lausanne
\\ 1015 Lausanne, Switzerland}
\corauth[cor1]{Corresponding author}

\begin{abstract}

Quantum efficiency studies for various wavelength and various
technical metal surfaces were carried out in a dedicated unbaked
vacuum chamber. Copper, magnesium, aluminium and aluminium-lithium
photocathodes were irradiated by two different high power, high
repetition rate, laser systems. We have observed an emission of
electrons for photon energy below the work function of the material.
This is explained by multiple photon absorption at the photocathode.
We have not observed any degradation of the QE for those materials,
but an improvement when irradiating them over a long period of time.
This is contrary to observations made in RF photoguns.

\end{abstract}


\journal{arxiv.org}

\end{frontmatter}


\section{Introduction}

In a free electron laser accelerator (FEL) one of the key component
is the electron source. The source should provide a sufficient
amount of electrons and should have a low emittance to provide the
x-ray photons requested by the end users. The electrons can be
produced by thermionic emission, photoemission or by field emission
\cite{ganter:2006,Togawa:2007,Ding:2009,Loehl:IPAC2010}. The
successful operation \cite{Emma:2010,Doumy:2011} of the first x-ray
FEL (XFEL) Linac Coherent Light Source (LCLS) using a Cu
photocathode has led the Paul Scherrer Institute (PSI) to adapt the
LCLS gun design for the future SwissFEL
\cite{Patterson:2010,SwissFEL:CDR}.

In the meantime, an injector test facility has been commissioned
with at its heart a RF photogun using a diamond milled oxygen free
electrolytic (OFE)Cu photocathode \cite{Schietinger:LINAC10}. The
electron production is insured by two different type of lasers, a
Nd:YLF (Jaguar$^{\registered}$ by Time-Bandwidth
\cite{Timebandwidth}) and a broadband Ti:Sa (Pulsar$^{\registered}$
by Amplitude Technologies\cite{AmplitudeTechnologies})
\cite{Vicario:FEL10}. In order to provide the required electron beam
quality (emittance) it is necessary to provide temporally \&
transversally shaped laser beam. This in turn has an impact on the
available laser energy at the cathode. Energy which is needed to
produce the amount of charge requested by the machine design
\cite{SwissFEL:CDR}. The amount of charges produced by the cathode
depends on its quantum efficiency (QE) (number of electron emitted
per number of incident photons).

In this technical note, we report on the QE of various metallic
photocathodes as a function of the laser wavelength. Cathode ageing
is often reported during RF photogun operation \cite{Lecce:2011}.
This ageing is characterized by a QE drop, which necessitates to
either repair, in-situ, the QE of the cathode or to exchange it
\cite{Lecce:2011}. We have then studied the evolution of the QE as a
function of the laser irradiation time for several metallic
cathodes.

\section{Experimental setup}

\begin{figure}[htbp]
\centering
\includegraphics[width=0.6\textwidth, clip=]{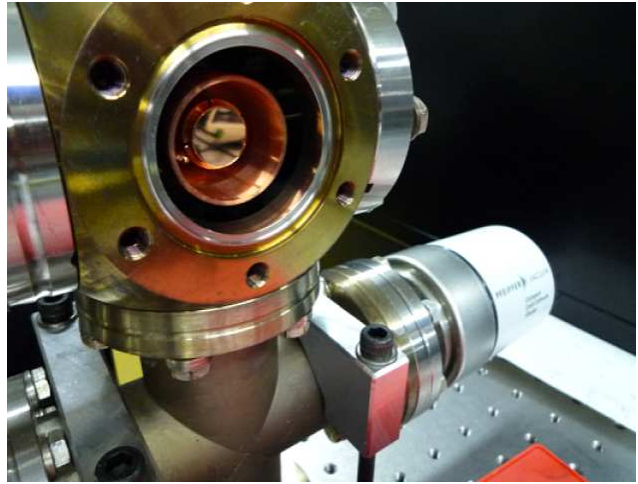}
    \caption{Experimental vacuum chamber with a Cu cathode visible at
    its center surrounded by a Faraday cup (FC). A cold cathode
    compact vacuum gauge monitors the total pressure in the chamber. A
    conflat MgF$_2$ vacuum window closes the vacuum
    chamber.}
\label{figExpSetupFOTO}
\end{figure}

The vacuum experimental system is shown in
Fig.\ref{figExpSetupFOTO}. The chamber is sealed using a transparent
conflat MgF$_2$ vacuum window, 90\% transmission from 200~nm to
3000~nm. The chamber is pumped through a 50~l/s turbo pump and is
not baked. The residual atmosphere is composed of water vapour for
more than 95\%. The other noticeable peaks are 2~uma, 28~uma and
44~uma. The vacuum is monitored by a compact cold cathode gauge. The
pressure during operation is between mid 10$^{-5}$~Torr at the start
of laser irradiation to mid 10$^{-7}$~Torr after a few days of
operation. The photocathode insert is placed at the center of the
chamber and it is surrounded by a Cu Faraday cup (FC).

\begin{figure}[htbp]
\centering
\includegraphics[width=0.6\textwidth, clip=]{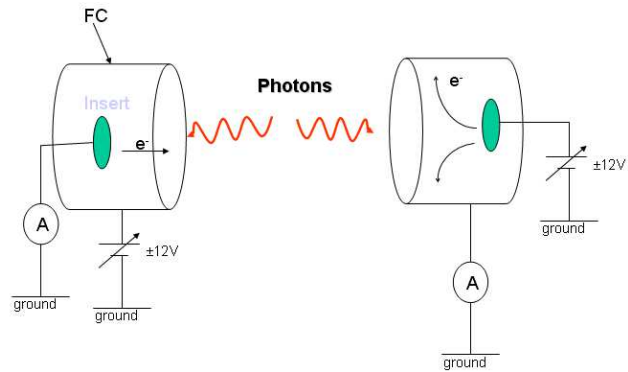}
    \caption{Experimental setup, Electric diagram.
    The system encompasses an insert (photocathode) a Faraday cup (FC) surrounding it, a +/- 12 V
    battery powered power supply, and a K6514 Ammeter }
\label{figExpSetup}
\end{figure}

The setup can be operated in two ways, shown in
Fig.\ref{figExpSetup}. Either by biasing the FC and recording the
current leaving the insert using a Keithley$^{\registered}$ K6514
Ammeter, or by biasing negatively the insert and recording the
current on the FC. A third option would have been to bias
negatively the insert and record the current through the K6514;
unfortunately the K6514 cannot be used in a floating mode.

We have used two types of lasers. A Ti:Sa working at 800~nm
wavelength, 100~fs pulse length with a 1~kHz repetition rate. The
800~nm is send to a Topas$^{\registered}$ \cite{Topas} optical
parametric amplifier (OPA) system which allows the selection of
various wavelength (UV to visible). The fluence used was
85.6~$\mu$J/cm$^2$ with a laser spot diameter between 2 to 3~mm. The
laser peak intensity with these parameters was 0.856~GW/cm$^2$.

The second laser is a Nd:YVO$_4$, Duetto$^{\registered}$ from
Time-Bandwidth Products \cite{Timebandwidth}, working at 355~nm
wavelength, with a 10~ps long pulse and a repetition rate set to
200~kHz. The laser average power has been varied from 115~mW to
300~mW depending on the sample. The typical laser spot diameter on
target was 8~mm. The laser fluence and peak intensity was,
respectively, 1.2~$\mu$J/cm$^2$ (0.12~MW/cm$^2$) and 3~$\mu$J/cm$^2$
(0.3~MW/cm$^2$). Both laser produce linearly polarized light.

\begin{figure}[htbp]
\begin{minipage}[t]{.5\linewidth}
\centering
\includegraphics[width=0.9\textwidth, clip=]{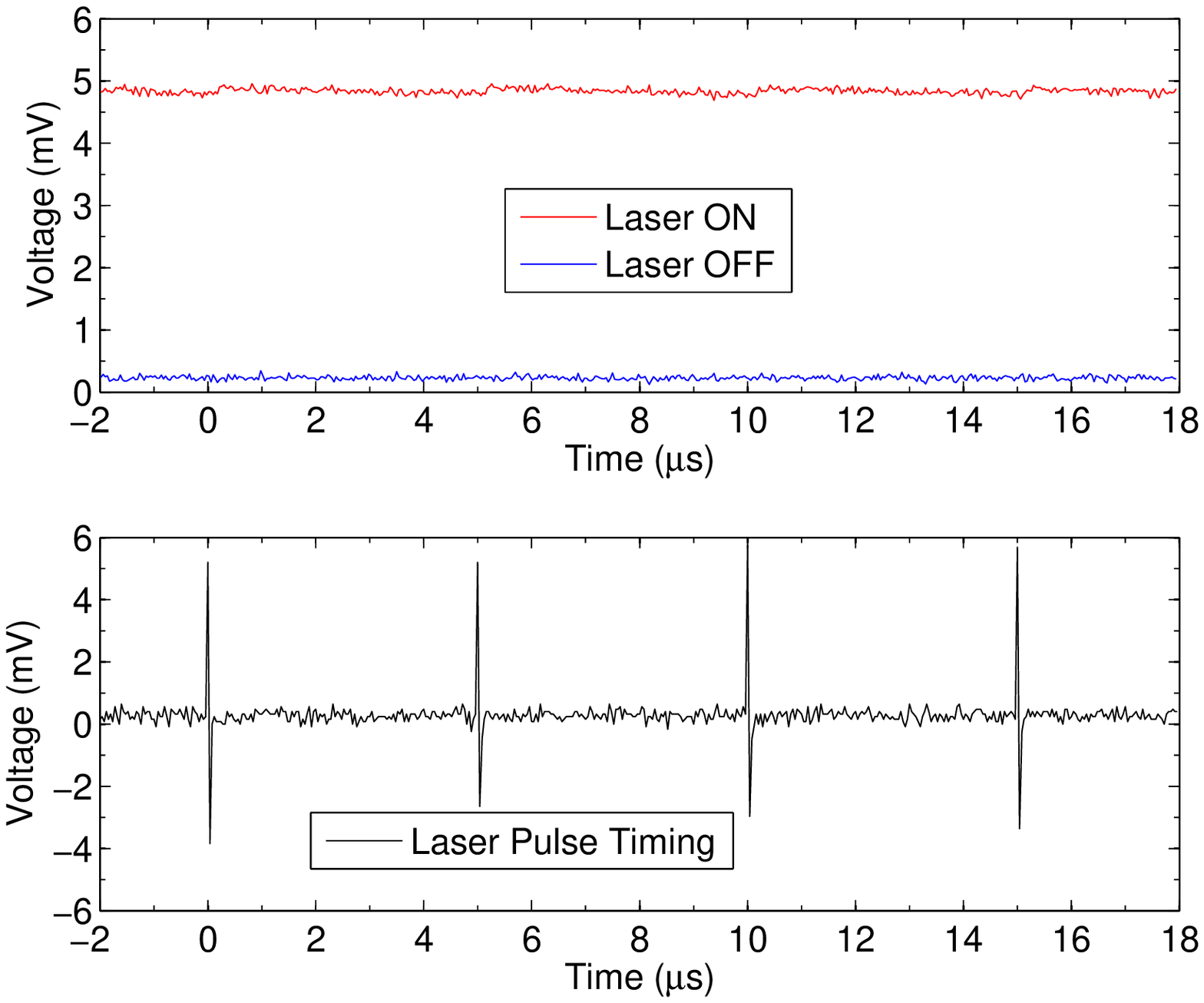}
\end{minipage}%
\begin{minipage}[t]{.5\linewidth}
\centering
\includegraphics[width=0.9\textwidth, clip=]{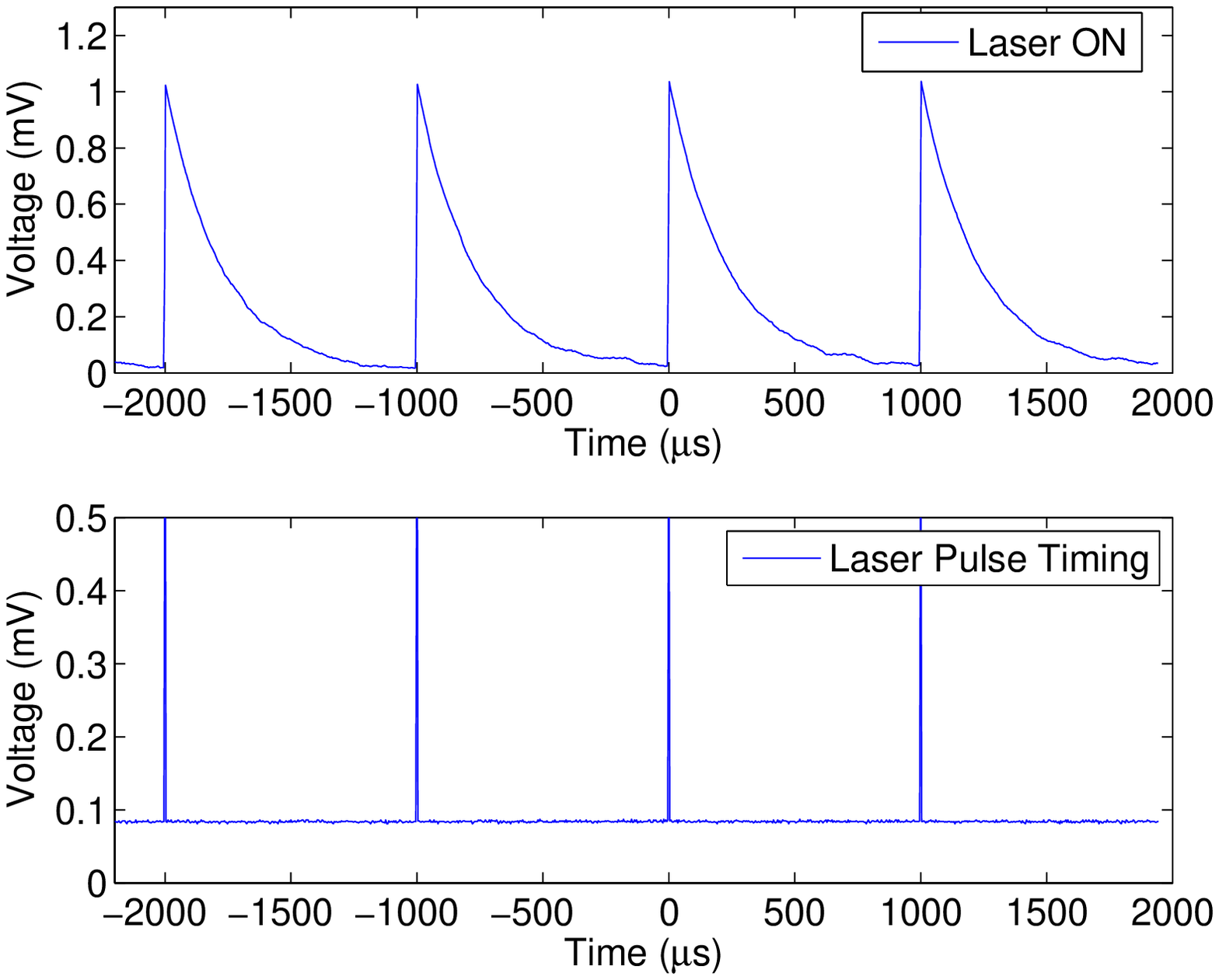}
\end{minipage}
\caption{The left figure (top) is the recorded voltage by the
oscilloscope when the Mg cathode is illuminated by the
Duetto$^{\registered}$ laser (355~nm, 200~kHz, 140~mW, $\sim$10~ps),
FC~=~+9~V. A mV corresponding to a nA of photo-emitted current. The
right figure (top) is the recorded voltage by the oscilloscope when
irradiating the same polished Mg cathode using the
Topas$^{\registered}$
    laser (260~nm, 1~kHz, 5~$\mu$J, $\sim$100~fs), FC is off.
    In both figures, the bottom plots are the traces of the recorded laser impulsion.}
\label{figOscilloRecord}
\end{figure}

In order to cross-calibrate the K6514 Ammeter, we have measured the
current coming from a freshly polished Mg cathode irradiated by the
Duetto$^{\registered}$ laser (355~nm; 200~kHz; 140~mW) with a
Tektronix-DPO 7254 oscilloscope, using the 1~M$\Omega$ entry
impedance. The same cathode, kept under primary vacuum for 3~months,
was also illuminated by the Topas$^{\registered}$ laser set to
(260~nm, 1~kHz, 5~$\mu$J), Fig.\ref{figOscilloRecord}.

The comparison shows excellent agreement between the K6514 reading
4.9~nA and the upper trace at 4.9~mV (laser ON) recorded by the
oscilloscope, left figure top plot in Fig.\ref{figOscilloRecord}.
This 4.9~mV equates to a current of 4.9~nA. The associated QE is
$\sim$0.023. The current read by the K6514 is similar, at the \%
level, when the FC power supply is switched off or when its voltage
is set to 0~V via its potentiometer. A zoom in of the laser ON trace
will show a small kick every 5~$\mu$s. This corresponds to the
200~kHz repetition rate pulse of the laser
(Fig.\ref{figOscilloRecord}, left figure bottom plot). The
oscilloscope cannot resolve the fast current signal produced by the
10~ps long laser pulse, due to the inherent impedance of the whole
system. This includes the capacitance of the BNC cable.
Consequently, the current displayed by the oscilloscope looks like a
DC offset. \newline In the case of the Topas laser, right plot, the
oscilloscope can resolve each pulse. The integrated signal (in nV.s)
for one pulse multiplied by the 1~kHz repetition rate is equal,
within 5\%, to the current recorded by the K6514.

The QE is the number of electrons emitted per irradiating number of
photons. The measurement system is not fast enough to resolve the
current at every pulse, like commonly achieved in an accelerator. We
hence define the QE by being the number of electrons coming from the
average current measured by the K6514 over the number of photons the
laser produces in one pulse. A laser delivering 1~mW with 1~kHz
repetition rate produces 1~$\mu$J/pulse, and with a 200~kHz
repetition rate, 5~nJ/pulse .

\section{Magnesium and Copper QE vs Wavelength}

The intrinsic emittance is the lower limit in beam emittance that
one can reach for a given cathode material, surface electric field
and laser wavelength. The intrinsic emittance (or thermal emittance)
can be expressed as follows, equation~\ref{EquThrmlEmitt}
\cite{Dowell:2009}:

\begin{eqnarray}
\varepsilon_{\rm thermal} = \sigma_x  \times \sqrt{\frac{h\nu -
\Phi_0 + e\ ^{3/2}.\sqrt{\frac{E}{4\pi\epsilon_0}}}{3m_0c^2}}
\label{EquThrmlEmitt}
\end{eqnarray}

Where the parameters are in SI unit : $\sigma_x$ the horizontal RMS
beam size, h$\nu$ the energy of the photons (J), $\Phi_0$ the work
function (WF) of a technical metal (J), which differs from an
atomically clean surface, $e$ the elementary charge, $E$ the applied
electric field (V/m), m$_0$ and $\epsilon_0$ the rest mass of the
electrons and the vacuum permittivity, respectively. In absence of
an external electric field, the best emittance is achieved when the
photon energy matches the WF of the element, unfortunately when this
condition is satisfied the QE drops to zero \cite{Dowell:2009}.

From the literature, one can find various WF for different clean
metals \cite{crc}. The WF of the metals is modified depending on
their surface chemistry and crystallographic orientation.
Table.\ref{tabWF} shows some of the WF for the bare metals we have
tested and the associated wavelength ($\lambda$).

\begin{table}[htbp]
\begin{center}
\caption{Work function and associated wavelength for some bulk
elements
\cite{crc,Chapman:1964,Assimos:1974,Koffyberg:1982,Brennan:2008}}
\begin{tabular}{|c|c|c|}
\hline Material& Work Function & Wavelength $\lambda$  \\
  & (eV) & (nm) \\
\hline Mg & 3.66 & 339  \\
\hline Al & 4.06 - 4.26 & 310 - 290  \\
\hline Cu & 4.53 - 5.10 & 274 - 245  \\
\hline MgO & 2.8 & 443  \\
\hline Al$_2$O$_3$ & 3.9 & 318  \\
\hline Cu$_2$O & 5.2 & 239  \\
\hline CuO & 5.3 & 234  \\
\hline Laser & 4.74 / 4.67 / 3.49 / 2.48 & 262 / 266 / 355 / 500 \\
\hline
\end{tabular}
\label{tabWF}
\end{center}
\end{table}

\subsection{QE vs Wavelength}

\begin{figure}[htbp]
\centering
\includegraphics[width=0.9\textwidth]{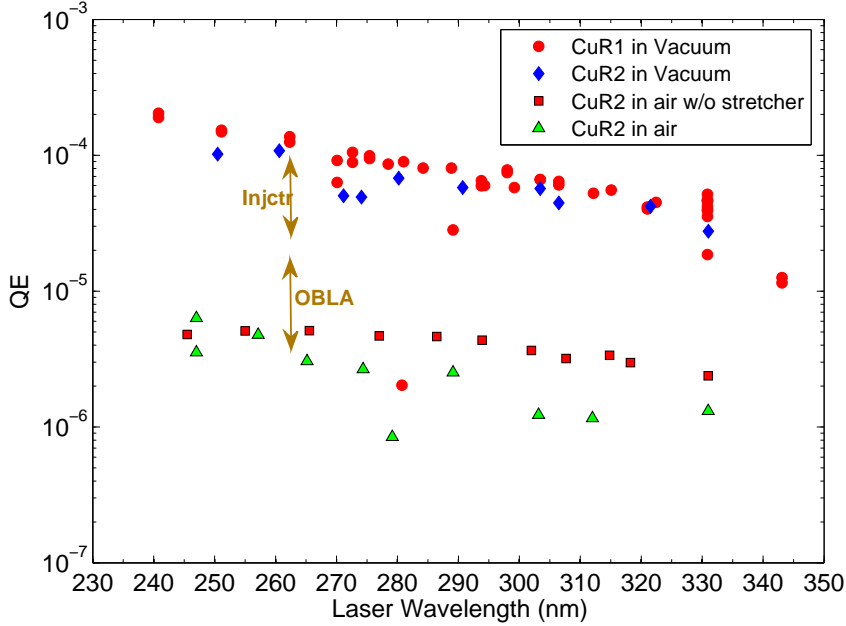}
     \caption{QE vs wavelength of two different polished (mirror-like) Cu samples. The FC is OFF.
     The QE are compared at a specific wavelength with QE data measured in two different accelerators.}
\label{figQECuR1R2}
\end{figure}

\begin{figure}[htbp]
\centering
\includegraphics[width=0.9\textwidth]{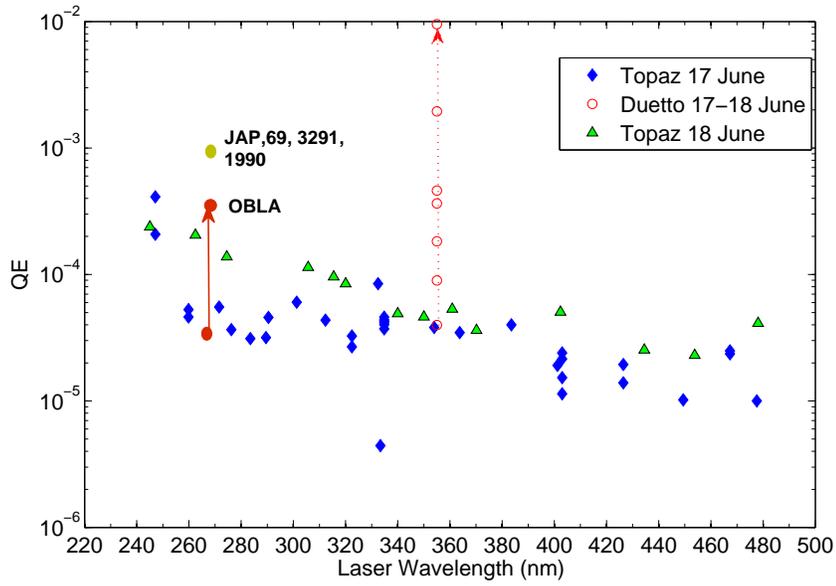}
     \caption{QE vs wavelength of polished (non mirror-like) Mg\#2. The FC is OFF.}
\label{figQEMg}
\end{figure}

We have measured the dependency of the QE as a function of
wavelength, provided by the Topas$^{\registered}$ OPA, for polished
and mirror-like, Cu Fig.\ref{figQECuR1R2}, and Mg Fig.\ref{figQEMg}.
According to the theory, we should see a sharp drop of the QE for
$h\nu-\Phi_0 \sim 0$. During the first experiment the FC was not
biased. The pressure during the experiment was in the low to mid
10$^{-6}$~Torr. The QE data are compared to QE measurements obtained
in a combined Diode-RF electron gun, labeled (OBLA) and using the RF
photogun of the SwissFEL injector, labeled (Injctr).


\begin{figure}[htbp]
\begin{minipage}[t]{.5\linewidth}
\centering
    \includegraphics[width=63mm,height=50mm,clip=]{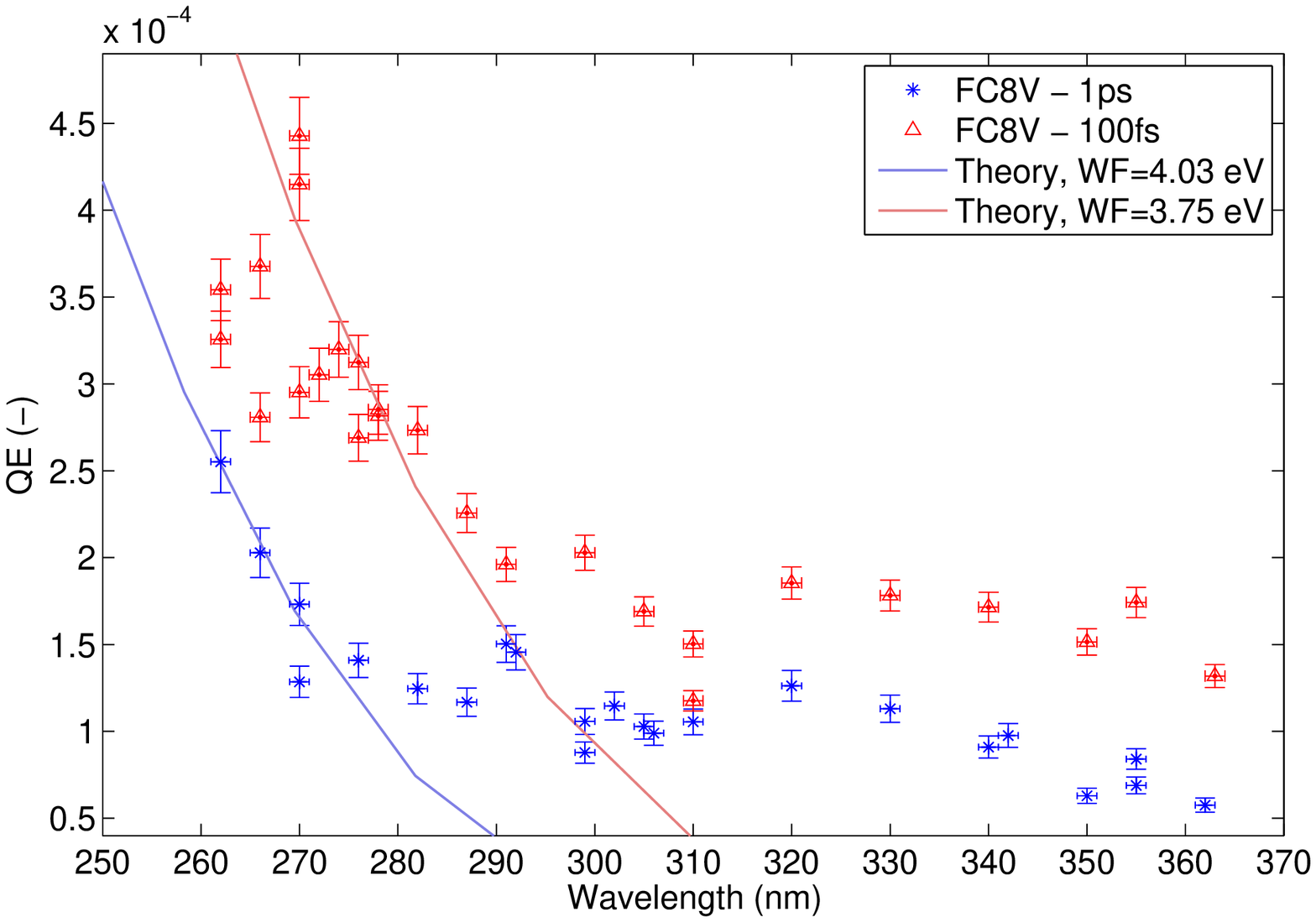}
\end{minipage}%
\begin{minipage}[t]{.5\linewidth}
\centering
    \includegraphics[width=63mm, height=50mm,clip=]{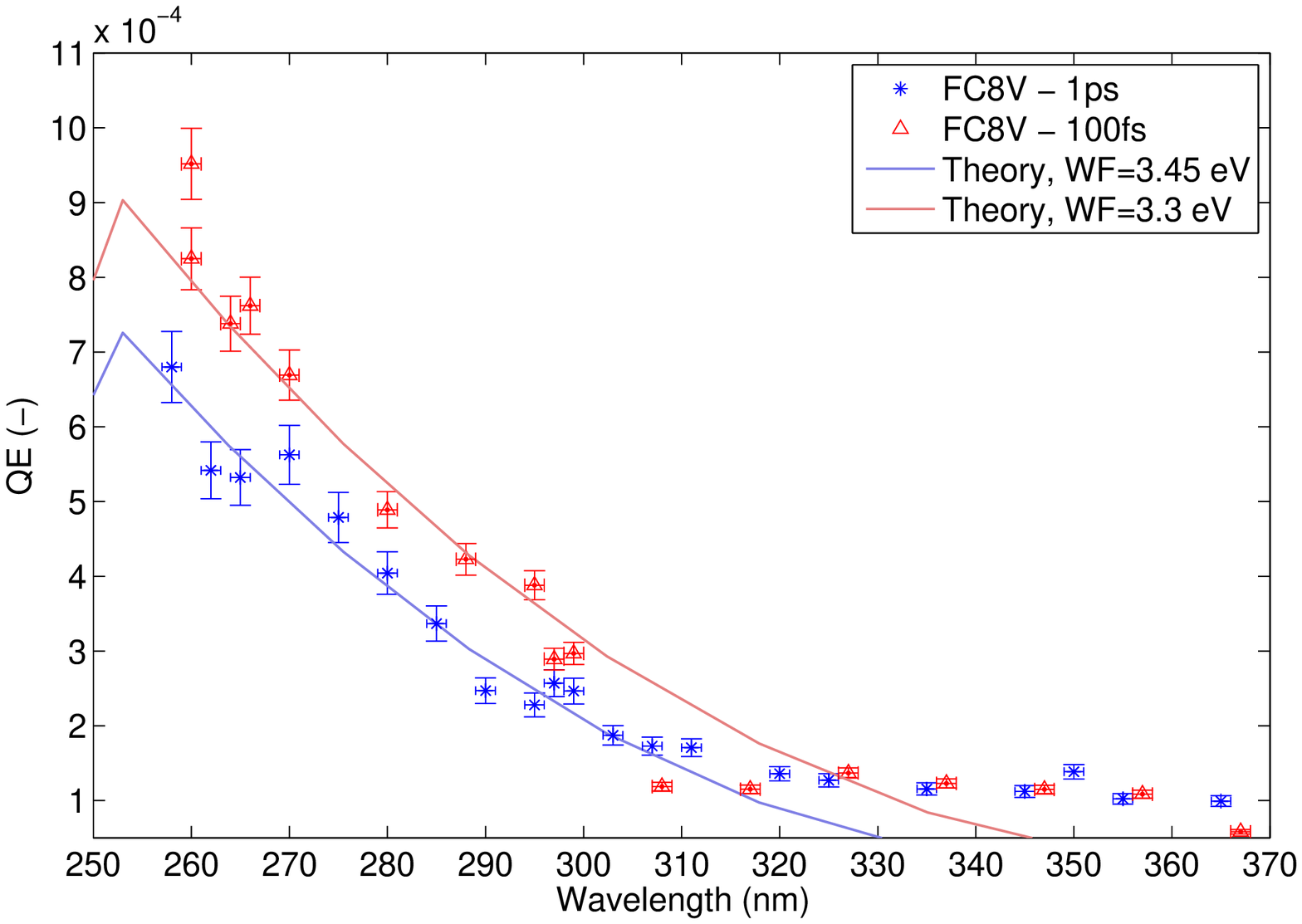}
\end{minipage}
    \caption{QE vs wavelength of a polished (mirror-like) Cu\#R4 sample (left plot)
    and of a freshly re-polished Mg\#2 (right plot), using two laser pulse lengths 1~ps and 100~fs.
    The straight line are the fittings using equation~\ref{EquQE}}
\label{figQECuR4Mg2}
\end{figure}

\begin{figure}[htbp]
\centering
\includegraphics[width=0.7\textwidth]{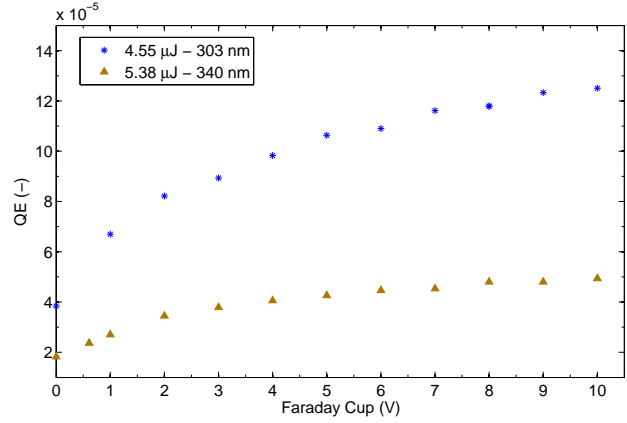}
     \caption{QE of Cu\#R4 vs the FC voltage at set wavelength
     for a set Topas$^{\registered}$ energy in microJ.}
\label{figQEvsFCCuR4}
\end{figure}

We have redone the experiment on Cu and Mg polished cathodes using
two different laser pulse length. The natural pulse length of the
Ti:Sapph is $\sim$100~fs. Using a glass stretcher the pulse length
is elongated to $\sim$1~ps. The results for the QE ($\lambda$)
presented in Fig.\ref{figQECuR4Mg2} have been obtained with the FC
set to +8~V. We have insured the collection of all electrons emitted
by setting properly the laser energy (in $\mu$J) at a given
wavelength (nm), as shown for example in Fig.\ref{figQEvsFCCuR4}.

\newpage
\subsection{Results above the WF}

Using the photoemission model described by Spicer \cite{Spicer:1969}
and utilizing equation~\ref{EquQE} for the QE($\lambda$)
\cite{Dowell:2009,Dowell:2006}, we have tried to fit the
measurements in Fig.\ref{figQECuR4Mg2}.

\begin{eqnarray}
QE(\omega) = \frac{1 - R(\omega)}{1 + \frac{\lambda_{opt}}{2 \
\lambda_{e-e}(E_m)} \ \frac{E_{ph}\sqrt{\Phi_{eff}}}{E_m^{3/2}}
(1+\sqrt{\frac{\Phi_{eff}}{E_{ph}}})} \times \nonumber \\
\frac{E_F +E_{ph}}{2E_{ph}} \times \left[1 - \sqrt{\frac{E_F +
\Phi_{eff}}{E_F + E_{ph}}}\ \right]^2
 \label{EquQE}
\end{eqnarray}

The validity of equation~\ref{EquQE} implies that $\frac{E_F +
\Phi_{eff}}{E_F + E_{ph}} < 1$, hence a photon energy above
$\Phi_{eff}$. The equation parameters are as follow. R is the
reflectivity, E$_F$ is the Fermi energy, E$_{ph}$ = $\hbar\omega$ is
the photon energy, $\lambda_{opt}$ is the laser penetration depth,
$\lambda_{e-e}$ is the electron-electron scattering length and E$_m$
is the energy above the Fermi level. $\Phi_{eff}$ is the effective
work function, which is the work function $\Phi_{0}$ of the bare
material minus the barrier reduction due to the external field
applied. In our case no strong electric field is present on the
cathode, hence $\Phi_{eff} \ (eV) \approx \Phi_{0}$.

The material reflectivity R, the optical penetration depth
$\lambda_{opt} = \frac{\lambda}{4 \pi k}$ are photon energy
dependent. $\lambda_{e-e}$ for Mg has been equaled to
$\lambda_{e-e}$ of Al. In the energy range considered for the
escaping electron the mean free path can vary very sharply
\cite{Spicer:1993sq,philiphofmann}.

The parameters used to fit the data for Cu and Mg are summarized in
Table.\ref{tabParameters}. For parameters varying in function of the
photon energy, a reference is given.

\begin{table}[htbp]
\begin{center}
\caption{Parameters used to fit the QE data for Cu and Mg of
Fig.\ref{figQECuR4Mg2} \cite{crc,Dowell:2006}}
\begin{tabular}{|c|c|c|}
\hline Parameters & Cu & Mg  \\
\hline R & \cite{crc} & \cite{crc}  \\
\hline E$_F$ & 7~eV  & 7.08~eV  \\
\hline $\lambda_{opt}$ & \cite{crc} (nm) & \cite{crc} (nm) \\
\hline $\lambda_{e-e}$ & 2.2~nm & 5~nm  \\
\hline E$_m$ & 8.6~eV & 8.6~eV \\
\hline $\Phi_{0}$ & 3.75 - 4.03~eV & 3.3 - 3.45~eV \\
\hline
\end{tabular}
\label{tabParameters}
\end{center}
\end{table}

The fits are rather insensitive to the value of $\lambda_{opt}$.
According to the above remarks on $\lambda_{e-e}$, the fits are
sensitive and the work functions $\Phi_{0}$ have to be adjusted by
0.1~eV. The reflectivity (R) is a very sensitive parameter. For Mg,
there is a gap for R between 3~eV and 5~eV photon energy \cite{crc}.
The fits shown in Fig.\ref{figQECuR4Mg2} (right plot) have been
obtained for a constant reflectivity (R=0.72). We have also
extrapolated the reflectivity values between 3 and 5~eV. The fits
using those value are no more correct. Fitting the data by utilizing
equation~\ref{EquQE} or the density of states of the material
\cite{CompES} will also show a discrepancy. This is explained as
equation~\ref{EquQE} is obtained by making use of the free-electron
gas model and by approximating the Fermi-Dirac distribution by an
Heaviside distribution \cite{Dowell:2009,Dowell:2006}.

Finally, The theory seems to fit relatively well the Mg
photoemission by assuming a work function which is between the pure
Mg metal and its oxide, Fig.\ref{figQECuR4Mg2} and
Table.\ref{tabWF}. For Cu, Fig.\ref{figQECuR4Mg2} (left plot), one
can fit well the data by using WF values below the clean Cu WF (up
to 1~eV, Table.\ref{tabWF}). The meaning of such fit could lead to
hypothesize that the Cu surface is contaminated by a chemical
elements like an hydride or an alkali. The Mg and Cu samples were
installed in the same manner using the same ethanol cleaned tools
and gloves. Contamination is unlikely to have occurred.

\subsection{Results below the WF}

We have observed that in all cases, the current recorded did not
drop to the noise level of the K6514 when the wavelength of the
laser was longer than the wavelength associated with the lowest WF
of the clean elements, see Table.\ref{tabWF}. Sub-ps or ps long
laser pulse seems to affect the surface so that photoemission is
possible for wavelength longer than the photoemission wavelength
threshold. When comparing the QE, for Mg or Cu, at pulse length of
100~fs and 1000~fs, The QE differs only a little,
Fig.\ref{figQECuR4Mg2}. Nevertheless it seems interesting to look at
different mechanism which could explain the electron emission at
wavelength above the material WF. Do we have :

\begin{itemize}
 \item[*] Thermal effects ?
 \item[*] Plasmonic effects ?
 \item[*] Multi-photon absorption ?
 \item[*] Mechanical effect (stress, strain) ?
 \item[*] Chemistry (oxide effect) ?
\end{itemize}

\subsubsection{Thermal effects}

\begin{figure}[htbp]
\begin{minipage}[t]{.5\linewidth}
\centering
\includegraphics[width=55mm,height=70mm,clip=]{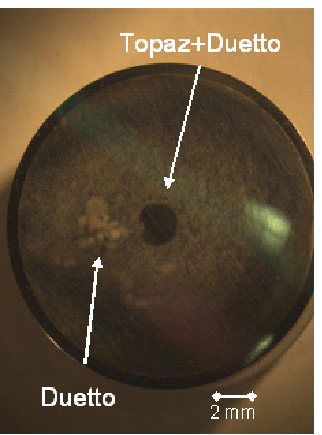}
     \caption{Damages caused by laser exposition on polished but rough Mg\# 2}
\label{figMgDamages}
\end{minipage}%
\begin{minipage}[t]{.5\linewidth}
\centering
\includegraphics[width=0.9\textwidth,clip=]{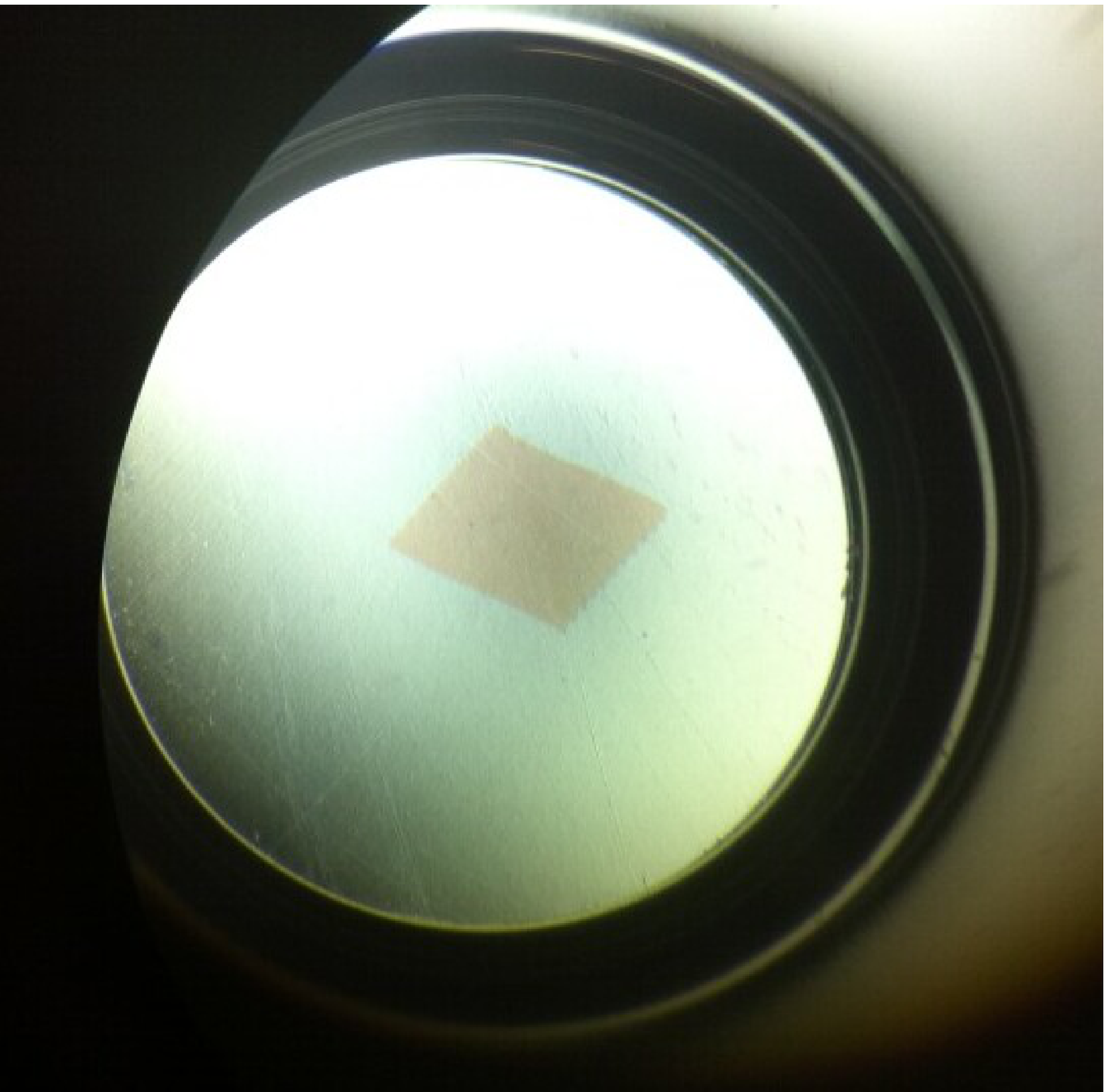}
     \caption{Damages caused by laser cleaning on polished Mg in the combined Diode-RF gun (OBLA) }
\label{figMgOBLADmgs}
\end{minipage}
\end{figure}

The increase of temperature due to the laser irradiating the
inserts can be calculated using equation~\ref{EquSpecificHeat}:

\begin{eqnarray}
Q = m \times C_s \times \Delta T
\label{EquSpecificHeat}
\end{eqnarray}

where Q is the heat energy (J), $C_s$ is the specific heat
(J/(kg.K)) and m the mass of the material (kg). We use at most
10$\mu$J of laser energy in the UV spectrum per pulse. We assume
that all the laser energy is converted into heat. We have also
considered, in Table.\ref{tabDeltaT}, that the UV light
($\sim$5.10$^{15}$~Hz) heats only a very thin layer of the insert
(skin depth) $\sim$1~nm thick, over the whole sample surface. If one
assumes that only the skin depth under the irradiated area of the
laser is heated then the temperature increase will exceed 900~K,
which is enough to melt Mg and Al. We have not observed any damages
on polished mirror-like samples. However, we have observed a dark
spot under an optical microscope for the Mg, Fig.\ref{figMgDamages}.
This black spot can also be observed on Mg cathodes which have
undergone laser cleaning in the combined Diode-RF electron gun
(OBLA) \cite{lepimpec:IPAC2010}, Fig.\ref{figMgOBLADmgs}. In the
OBLA experiment, we have rastered the laser beam on the Mg surface.
The laser fluence could be up to 40~mJ/cm$^2$ at 266~nm, using a
10~ps long laser pulse with a repetition rate of 10~Hz. The diamond
area cleaned, Fig.\ref{figMgOBLADmgs}, was obtained using
12~mJ/cm$^2$ of laser fluence. In the small chamber experiment, the
fluence of the Topas$^{\registered}$ and Duetto$^{\registered}$ on
Mg\#2 were respectively less than 85~$\mu$J/cm$^2$ (2-3~mm laser
spot size) and 1.2~$\mu$J/cm$^2$ (8~mm laser spot size).

\begin{table}[htbp]
\begin{center}
\caption{Temperature increase of different metals when submitted
to 10~$\mu$J/pulse laser irradiation.}
\begin{tabular}{|c|c|c|c|c|}
\hline Metals & Mass  & Specific Heat & $\Delta$T & $\Delta$T ($\sim$1~nm)  \\
  & (kg) & (J/(kg.K)) & (K) & (K)  \\
\hline Cu & 0.0028 & 387 &  9.2 10$^{-6}$ & 19 \\
\hline Mg & 0.0021 & 1050 &  4.5 10$^{-6}$ & 23 \\
\hline Al & 0.0014 & 900 &  7.9 10$^{-6}$ & 42\\
\hline
\end{tabular}
\label{tabDeltaT}
\end{center}
\end{table}

From the consideration above, we think that thermal effects cannot
account for the emission of light for laser energy  below the work
function.

\subsubsection{Plasmon assisted photoemission}

Surface plasmon can enhance the photoemission of surfaces
\cite{Sipe:1981,Tsang:1990,Tuske:1997,Guo:2010}. The photoemission
enhancement occurs for some angle of incidence and is also
polarization dependent. We operate at normal incidence, and both
lasers are linearly polarized in the plane of the photocathode. At
normal incidence the probability of plasmon excitation is usually
quasi null \cite{Tsang:1990}. However, it seems that when the
roughness of the surface is on the order of a few 10's of
nanometers, surface plasmon could be excited hence enhancing
photoemission \cite{Qian:2010}.

\subsubsection{Mechanical effect (stress, strain)}

All of our inserts are mechanically polished. This induces
stresses in the metal and on its surfaces. After polishing, the
inserts are not baked, hence not stress relieved. Mechanical
stress has been shown to modify the WF of metals \cite{Li:2004}.
However, the modification seems not to be sufficient to account
for the photoemission at photon energy below 4.1~eV.

\subsubsection{Chemistry on the surface}

The presence of oxygen on the surface can significantly modify the
the WF of a bare metal, either by lowering it or by increasing it
\cite{Chapman:1964,Anderson:1959,Grubb:2009}. Although contaminant
layers produced by air exposure are usually detrimental to the QE
\cite{Lecce:2011,Lorusso:2011}, applying a thin film of MgO over
silver can not only lower the Ag work function but may also be
beneficial for the electron beam emittance in an accelerator
\cite{Nemeth:2010}.

We believe that some part of the photoemission curve for Mg,
Fig.\ref{figQEMg}, can be explained by the presence of the oxide. On
the contrary natural copper oxide seems to have a work function
above 5~eV \cite{Assimos:1974}, hence a detrimental oxygen chemistry
compared to Mg.

\subsubsection{Multiphoton absorption}

\begin{figure}[htbp]
\centering
\includegraphics[width=0.8\textwidth,clip=]{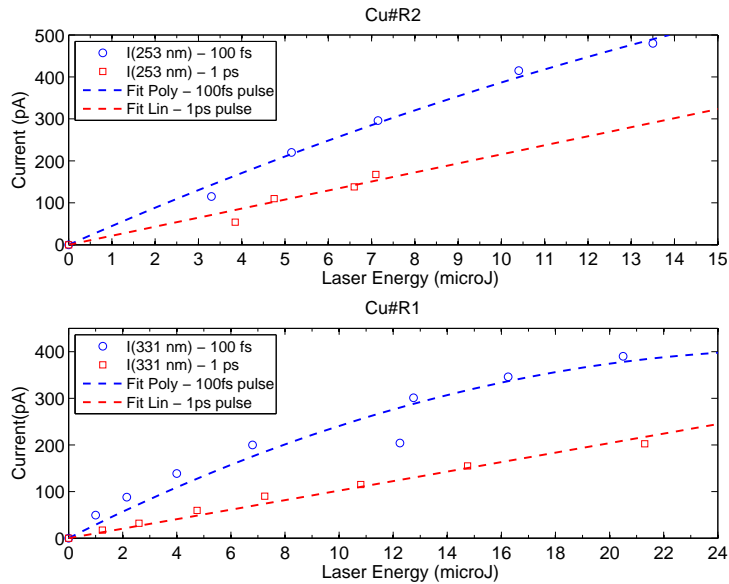}
     \caption{Electron Current extracted versus the Topas$^{\registered}$ laser energy
     at 253~nm for CuR2 and at 331~nm for CuR1. In both cases the FC is OFF.}
\label{figxxnmR1R2}
\end{figure}

Multi-photon absorption, mainly two-photon absorption, is possible
if the density of photons impinging the surface is high enough.
Both laser Topas$^{\registered}$ and Duetto$^{\registered}$ can
deliver such intensities. Fig.\ref{figxxnmR1R2} shows the
intensity of current extracted from Cu at 253~nm and at 331~nm for
Topas$^{\registered}$ pulses of 100~fs and 1~ps.

For photon energy above the WF, like for $\lambda$=253~nm, one
should expect to see a linear dependence of the current vs the laser
energy. For photon energy below the WF, for $\lambda$=331~ nm, one
should see a quadratic dependence of the current vs the laser
energy. For Cu, for both wavelength, we have a linear dependency of
the current emission versus the laser energy for laser pulse length
of 1~ps. For laser pulse length of 100~fs, the dependency seems
quadratic. However, the quadratic dependence should have a positive
bending and not a negative one as shown. This quadratic negative
bending dependence is often seen in RF photogun when the bunch space
charge hampers the emission of electrons \cite{Lecce:2011}. A linear
fit through zero can also be satisfactorily applied for Cu\#R2
irradiated at 253~nm with a 100~fs laser pulse length. This fit
implies an electron emission by a one-photon photoelectric effect
which is non space charge limited.

For Mg\#2, we have measured the current emission vs the laser
energy. For $\lambda$=247~nm the dependance is linear. For the
following wavelength $\lambda$ (403~nm, 520~nm) the dependance is
quadratic with a positive curve, hence implying a photoelectric
effect driven by a double-photon absorption.

\subsection{Comparison with other data}

We have compared the data obtained in this chamber with data
obtained inside an RF photogun. For Cu it is usual to find values
for the QE in the 10$^{-5}$ range. We measured similar values in our
combined Diode-RF gun, labeled OBLA in Fig.\ref{figQECuR1R2}, and in
our RF photoinjector, labeled Injctr in Fig.\ref{figQECuR1R2}. At
250~nm, QE value of 1.4~10$^{-4}$ is also reported. This is
consistent with our measurements
\cite{Lecce:2011,Dowell:2010,Tkachenko:2010}.

We have also reported some of the QE data obtained on bulk or thin
film Mg obtained in our combined Diode-RF gun, labeled OBLA. The
results obtained are compatible with literature data
\cite{Lecce:2011,Qian:2010,Lorusso:2011,Rao:1990}.

\section{Duetto laser cleaning}

\subsection{Cu and Mg}

\begin{figure}[htbp]
\begin{minipage}[t]{.5\linewidth}
\centering
\includegraphics[width=0.9\textwidth,clip=]{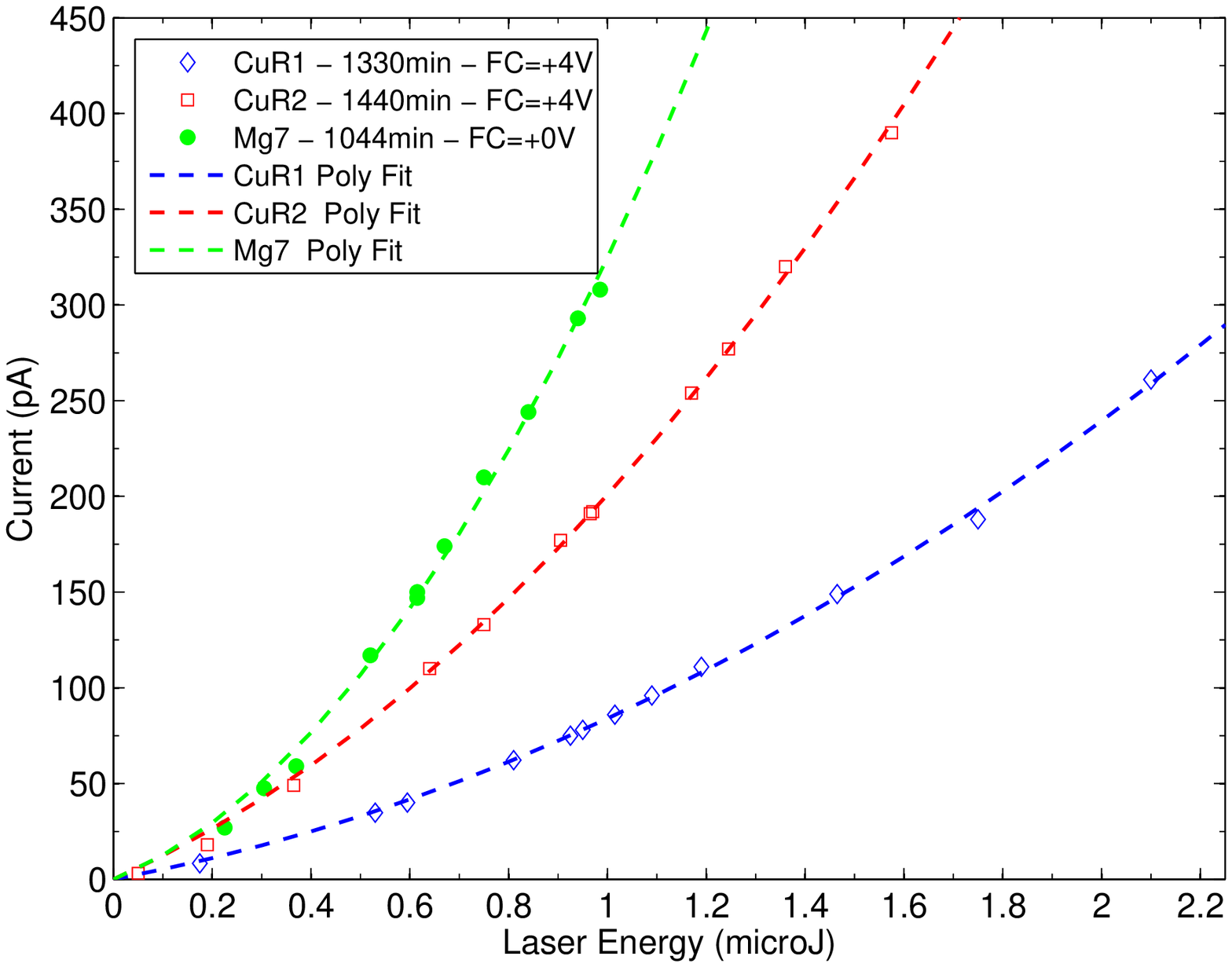}
\end{minipage}%
\begin{minipage}[t]{.5\linewidth}
\centering
\includegraphics[width=0.9\textwidth,clip=]{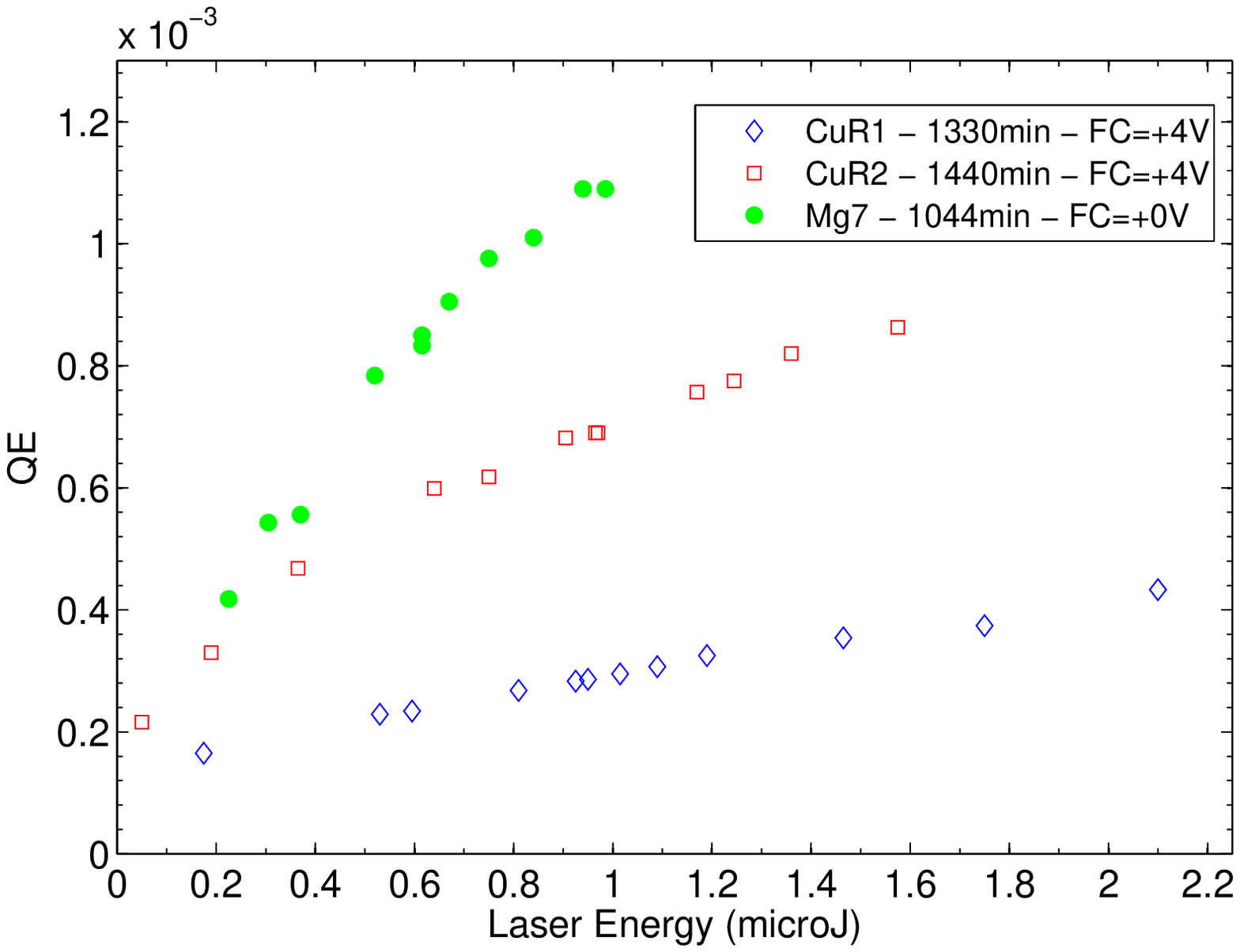}
\end{minipage}
     \caption{Current (left) and QE (right) measured as a function of the Duetto$^{\registered}$
     laser energy for two Cu samples and one Mg insert.
     Double-Photon absorption emission is responsible for electron emission
     For Cu the FC is on at +4~V and is off for the Mg insert.}
\label{figIvsLasECuMg}
\end{figure}

\begin{figure}[htbp]
\centering
\includegraphics[width=6.5cm, height=10cm,angle=-90,clip=]{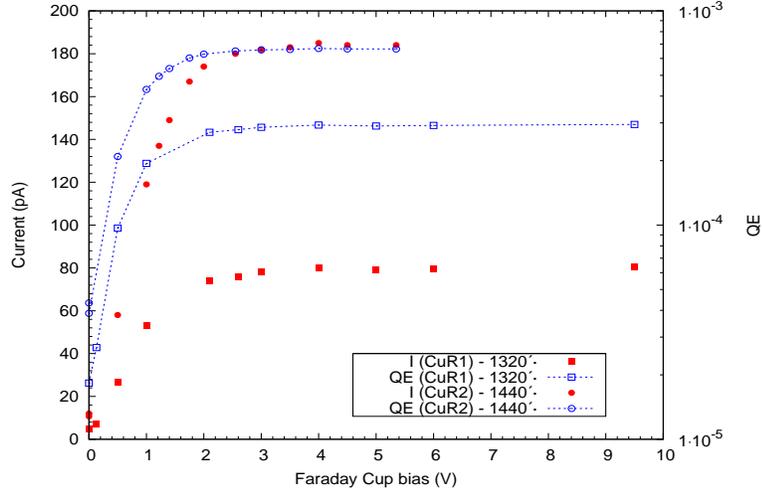}
     \caption{QE and Current versus the FC bias of two polished mirror-like Cu samples  }
\label{figCuR1R2FC}
\end{figure}

\begin{figure}[htbp]
\centering
\includegraphics[width=0.7\textwidth,clip=]{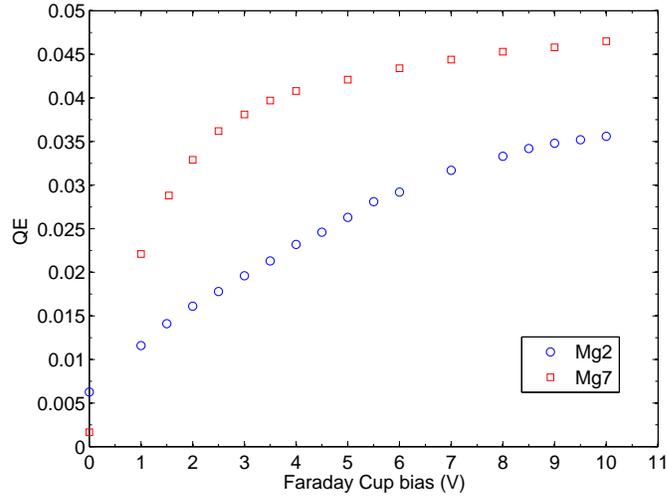}
     \caption{QE versus the FC bias of two polished Mg sample Mg\# 2 (rough) and Mg \# 7
     (mirror-like)}
\label{figMg27FC}
\end{figure}

At 355~nm we have measured the QE of Mg\#2 using the
Duetto$^{\registered}$ laser, Fig.\ref{figQEMg} (circles). The
chamber was partially vented to 0.1~mbar of air atmosphere during
the transfer of the chamber from one location to another. The
initial QE is similar to the QE measured with the
Topas$^{\registered}$ laser. The QE increases with the irradiation
time. The laser spot size was chosen to be 8~mm in diameter on the
cathode with a fluence of usually 1.2~$\mu$J/cm$^2$. The fluence per
pulse is $\sim$80 times lower that when using the
Topas$^{\registered}$. Fig.\ref{figIvsLasECuMg} (left) shows the
current extracted from two polished Cu cathodes, which were HCl
cleaned to remove the oxide layer and then rinsed with alcohol
before mounting. Fig.\ref{figIvsLasECuMg} also shows the current
extracted from a polished and mirror-like Mg photocathode (Mg\#7)
for comparison. The laser peak intensity, associated to the laser
energy, varies from 0 to 0.45 MW.cm$^{-2}$. The QE in function of
the laser energy is shown on the right plot of
Fig.\ref{figIvsLasECuMg}. For the Cu cathodes, the data plotted were
measured while the FC was set to + 4~V and 0~V for Mg\#7 cathode.
Fig.\ref{figCuR1R2FC} and Fig.\ref{figMg27FC} show the effect on the
current/QE measured when applying a small positive voltage on the
FC.


The electron current measured while irradiating at 355~nm
wavelength, 10~ps pulse length, is due (for Cu and Mg) to the
double-photon absorption, square fitting of the current vs the laser
energy in Fig.\ref{figIvsLasECuMg}. We have seen no black marking
like in Fig.\ref{figMgDamages}, on either Cu or Mg samples. The
samples stayed pristine after a few days of irradiation. The
conditioning in time of Cu was carried out with the
Duetto$^{\registered}$ laser delivering 200~mW of power, while being
130~mW for Mg\#7.

As already mentioned, Fig.\ref{figCuR1R2FC} and Fig.\ref{figMg27FC}
show the effect on the current/QE measured when applying a small
positive voltage on the FC. The increase of current extracted is
usually above a factor~10 when applying a few Volts on the FC.


The total vacuum pressure of the unbaked vacuum system during the
laser exposure (2300~min) dropped from 5.10$^{-5}$~Torr to
5.10$^{-6}$~Torr. The evolution of the QE for Cu, goes from the mid
10$^{-6}$ range to the mid 10$^{-5}$ range when the FC power supply
is turned off, data not shown. The QE, FC ON, for Cu is $\sim$0.1\%
(Fig.\ref{figCuR1R2FC}) and $\sim$4.5\% for Mg
(Fig.\ref{figMg27FC}). Those QE are much higher than the QE measured
for photocathodes installed in an RF photogun. This is explained by
the fact that in an accelerator the charge per bunch, hence per
laser pulse, is accurately measured, while here we measure the
average current.

The volume of interaction between the residual gas and the laser
beam is of a few cm$^3$. The number of molecules at 10$^{-6}$~Torr
and in a cm$^3$ is in the order of 3.10$^{10}$ and the number of
photons in the beam is $\sim$ 3.10$^{12}$. The usual cross section
of interaction is 10$^{-16}$~cm$^2$. This would amount to a pA of
current, if all ions were falling on the photocathode. The main
effect of the FC bias is the suppression of the space charge present
at the cathode surface hampering the emission of electrons. This
space charge is weak enough that only a few Volts are sufficient to
counterbalance it, as shown by the current plateau for bias above
+4~V, Fig.\ref{figCuR1R2FC}.

\begin{figure}[htbp]
\centering
\includegraphics[width=0.8\textwidth]{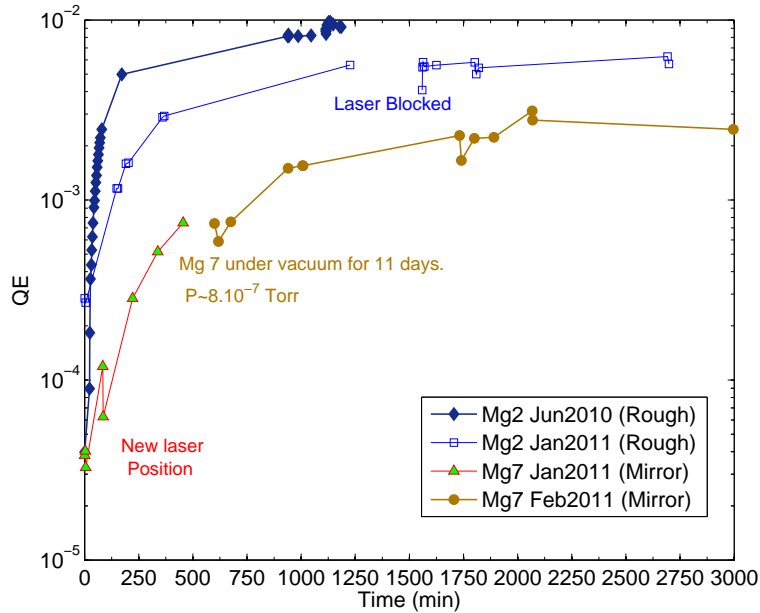}
     \caption{QE evolution under Duetto irradiation for two polished
     Mg sample Mg\#2 (rough) and Mg\#7 (mirror-like). FC is off in all cases.}
\label{figMg27History}
\end{figure}

We have also monitored the evolution in time of the QE for Mg
cathodes when irradiated by the Duetto$^{\registered}$ at 355~nm,
Fig.\ref{figMg27History}. The laser size on cathode was 8~mm and the
repetition rate 200~kHz for all the data plotted. The full diamond
plot labelled "Mg2 Jun2010" is the same data plotted in
Fig.\ref{figQEMg} (open circles). The laser cathode cleaning with
low energy per pulse increases the QE by a few order of magnitude
after less than 10~hours of exposure. The laser was blocked for one
hour and the system kept in vacuum. Upon restart the QE barely
dropped. The same cathode was reused 6~months later. The cathode was
left in the vacuum chamber, which was air-vented and the vacuum
valve was closed. After pump down, the laser was shined on the
cathode with the same parameters as used previously, including an
average power of $\sim$130~mW; open square data labelled "Mg2
Jan2011". The QE almost reached the same value as before in the same
amount of time. The laser was blocked over night, with the vacuum
pressure improving during the night. Again the drop in QE was
minimal upon resuming irradiation. Damages, on the polished but
non-mirror like sample, from the irradiation by both lasers have
been shown in the left photo of Fig.\ref{figMgDamages}.

A freshly polished Mg (Mg\#7) insert was installed in the vacuum
chamber, and exposure started when the surrounding pressure was
$\sim$5.10$^{-5}$~Torr. The parameters used for the irradiation were
similar to the one used for sample Mg\#2. The first QE value
recorded was lower than for a rougher surface. This is not
surprising as rougher surfaces will trap more photons, hence
increasing the production of electrons compared to a flatter
surface. The QE of the mirror-like surface stays systematically
lower than for a rougher surface.

After a couple of hours of irradiation on Mg\#7,
Fig.\ref{figMg27History} (full triangle), the laser spot was moved
to another location on the cathode. The sample size was 14~mm and
the spot size was 8~mm. It is possible that some overlapping of the
already conditioned area occurred. This would explain why the QE
reduction seen does not fall at the starting value. After a few more
hours of laser exposure, the laser was taken away and the system
kept in a dynamic vacuum for 11~days. After re-start of the
illumination (full circle), the QE value of Mg\#7 is similar to the
end-of-illumination QE value (full triangle),
Fig.\ref{figMg27History} .
\newline It seems to be commonly acknowledged that long exposure to the
vacuum residual gas is detrimental to the QE \cite{Lorusso:2011}.
This result on Mg and some more on Cu (Fig.\ref{figTopasCuR2}) and
AlLi (Fig.\ref{figQEvsFCAlLi}) are in flagrant contradiction.

Using a high repetition rate, short pulse laser, with a wavelength
longer than the associated wavelength of the WF, at a low fluence
(less than a microJ of energy per pulse over a "broad" area), in an
unbaked vacuum atmosphere has proven efficient in increasing the QE
of both Cu and Mg cathodes. In both cases multi-photon absorption is
responsible for electron emission. In the case of Mg, Mg oxide can
have a lower work function that pure Mg and the emission at 355~nm
can be normal photoemission instead of double-photon absorption.
After a few hours of laser exposure, we have not seen any linear
dependence of the current extracted versus the laser energy.

\subsection{Aluminium and Aluminium Lithium alloy}

Similarly to what was done on copper and magnesium, we have
irradiated an aluminium (Al) sample and an Aluminium Lithium
(AlLi) alloy insert.

Al and AlLi photocathode could be alternatives to Mg as a
photocathode. In the Diode-RF electron gun (OBLA) we have measured
the emittance and the QE of various metals
\cite{Lecce:2011,lepimpec:IPAC2010,Hauri:2010}. Respective to the
QE, Aluminium has been found better than Copper and not as good as
Magnesium. The emittance was higher than for Cu. According to
literature Mg also produces lower emittance than Cu
\cite{Qian:2010}. What could make Al still attractive, is its higher
vapour pressure upon baking compared to Mg. Most photo-RF guns are
baked before RF processing is started, although the temperature
might be less than 150~C for a long period of time.

As an alternative to pure Al is AlLi alloy \cite{Septier:1992}.
Lithium and Magnesium are in the first and second group of the
periodic table. They are both strongly reactive to oxygen. Both
elements might migrate on top of the surface of Al over time
\cite{lepimpec:2005} which might be the reason for the enhanced
production of electrons compare to the bare Al.

Al$_{95}$Li$_{2.5}$Cu$_{1.5}$Mg$_1$ alloy was bought from
Goodfellow$^{\registered}$ in tube form. The tube was smashed into a
circular insert then polished (mirror surface like) and kept in air
for 3~months before installation.

\begin{figure}[htbp]
\centering
\includegraphics[width=6.5cm, height=10cm,angle=-90,clip=]{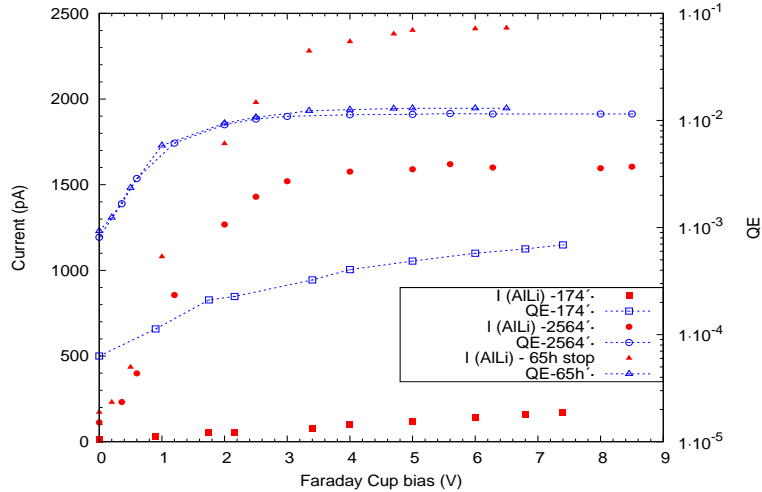}
     \caption{QE and Current evolution of AlLi alloy during Duetto$^{\registered}$ exposition,
     in function of the FC bias voltage.}
\label{figQEvsFCAlLi}
\end{figure}

The irradiation with the Duetto$^{\registered}$ laser produced only
a few pA of currents, independently of the FC voltage) even when
using 380~mW of laser power, fluence~3.8$\mu$J/cm$^2$. The current
extracted did decrease from 3~pA to 1~pA in 20~min. The sample was
sent to polishing and re-installed. The results of the current and
QE obtained at different times of exposure versus the FC voltage are
shown in Fig.\ref{figQEvsFCAlLi}.

\begin{figure}[htbp]
\centering
\includegraphics[width=0.7\textwidth, clip=]{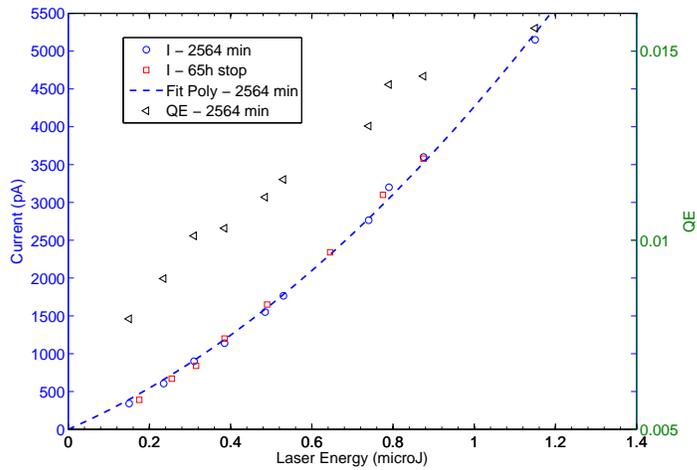}
     \caption{AlLi current extracted in function of the Duetto$^{\registered}$ laser energy.
     FC is set at +4V. }
\label{figIvsLasEAlLi}
\end{figure}

The production of electrons is done through the double-photon
absorption process as seen by the square dependency fit in
Fig.\ref{figIvsLasEAlLi}. The data labelled "65h stop" on both
figures, Fig.\ref{figQEvsFCAlLi} and Fig.\ref{figIvsLasEAlLi}, have
been obtained at re-start of irradiation after 65~hours of laser
downtime. During that time the chamber stayed actively evacuated by
the vacuum pump system. As for Cu or Mg, the QE of the AlLi did not
decrease while the sample stayed in an unbaked vacuum (P$\sim$2.6
10$^{-6}$~Torr) for a few days.

For extracted current above 1500~pA, during the QE vs FC voltage
scan, and after every step up of the FC voltage the current reads
high and can drop by 100~pA in 30~s. For extracted current below
1500~pA, the current drops by a few pA. After this initial drop in
extracted current, the charge extracted increases again slowly under
the Duetto$^{\registered}$ irradiation. This behaviour, initial
extracted current drop after an increase of laser power, was not
observed on either Cu or Mg.

\begin{figure}[htbp]
\centering
\includegraphics[width=6.5cm, height=10cm,angle=-90,clip=]{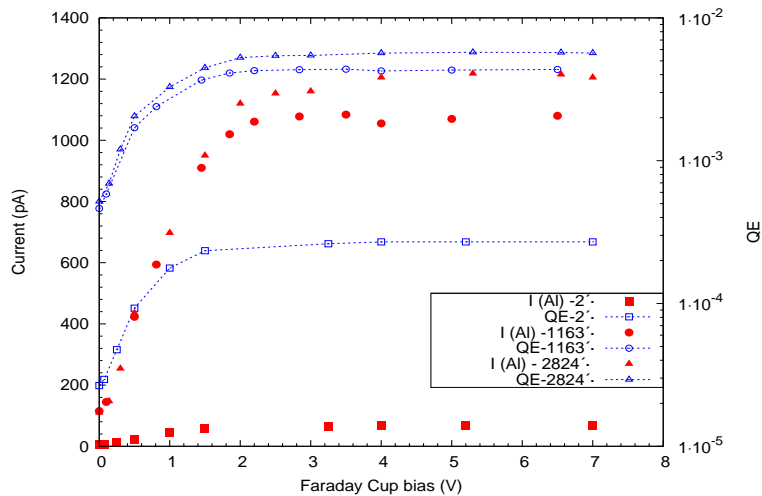}
     \caption{QE and Current evolution of Al during Duetto$^{\registered}$ exposition,
     in function of the FC bias voltage.}
\label{figQEvsFCAl}
\end{figure}

Finally we tested a freshly polished Al sample. The QE and current
extracted versus the FC bias is shown in Fig.\ref{figQEvsFCAl}, at
different times of laser exposure. A previous attempt of measuring
the QE vs wavelength on an Al sample kept in air for a long time did
not give consistent results as were obtained for Cu or Mg,
Fig.\ref{figQECuR1R2} and Fig.\ref{figQEMg}.

Al and AlLi alloy reacted similarly in terms of QE. The initial QE,
FC OFF, is in the 10$^{-5}$ range and increases by a decade at the
end of the laser exposure. When turning on the FC, the end value for
the QE (for AL and AlLi) is close to a percent. In both cases the
photoemission is double-photon absorption. It seems that for
Aluminium-based photocathode, a freshly prepared cathode behaves
better than a cathode well prepared and then kept in air for a few
months. This was not seen for Mg or Cu. No damage was seen on either
Al or AlLi sample after removal from the chamber.

\section{Ageing of Cu}

LCLS has shown that a high current, XFEL operation compatible
emittance, can be obtained from a well-prepared Cu photocathode. A
long lifetime between cathode exchange is also beneficiary for user
operation. However, as the cathode ages the QE drops and in-situ
techniques, like laser cleaning, can be applied to regenerate the
QE. On the right photograph of Fig.\ref{figMgDamages}, we have seen
how an aggressive laser cleaning can modify the surface. Laser
induced surface alteration was also seen on the Cu cathode of LCLS
or in the SPARC experiment \cite{Lecce:2011}.

\begin{figure}[htbp]
\begin{minipage}[t]{.5\linewidth}
\centering
\includegraphics[width=0.92\textwidth,clip=]{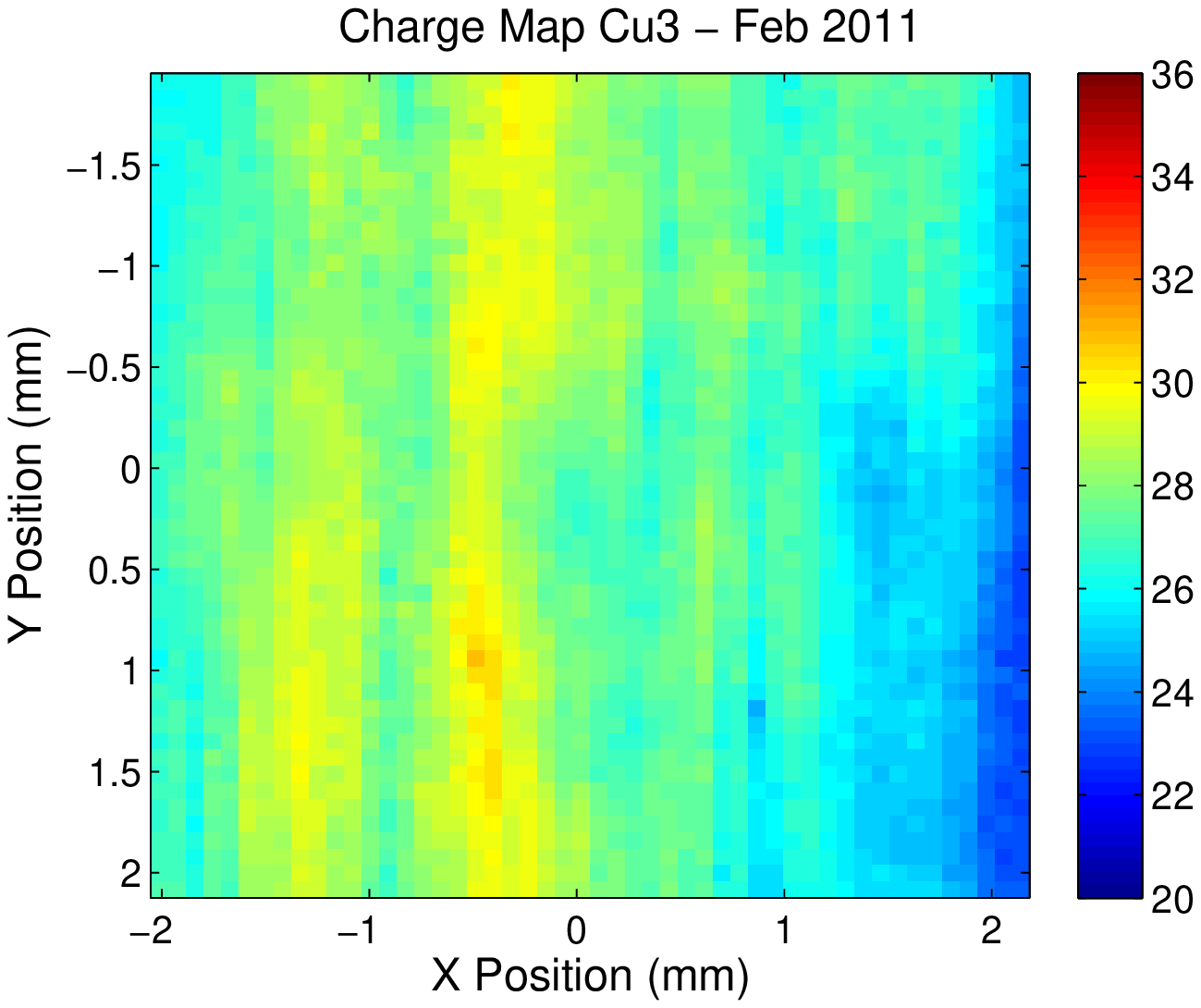}
\end{minipage}%
\begin{minipage}[t]{.5\linewidth}
\centering
\includegraphics[width=0.92\textwidth,clip=]{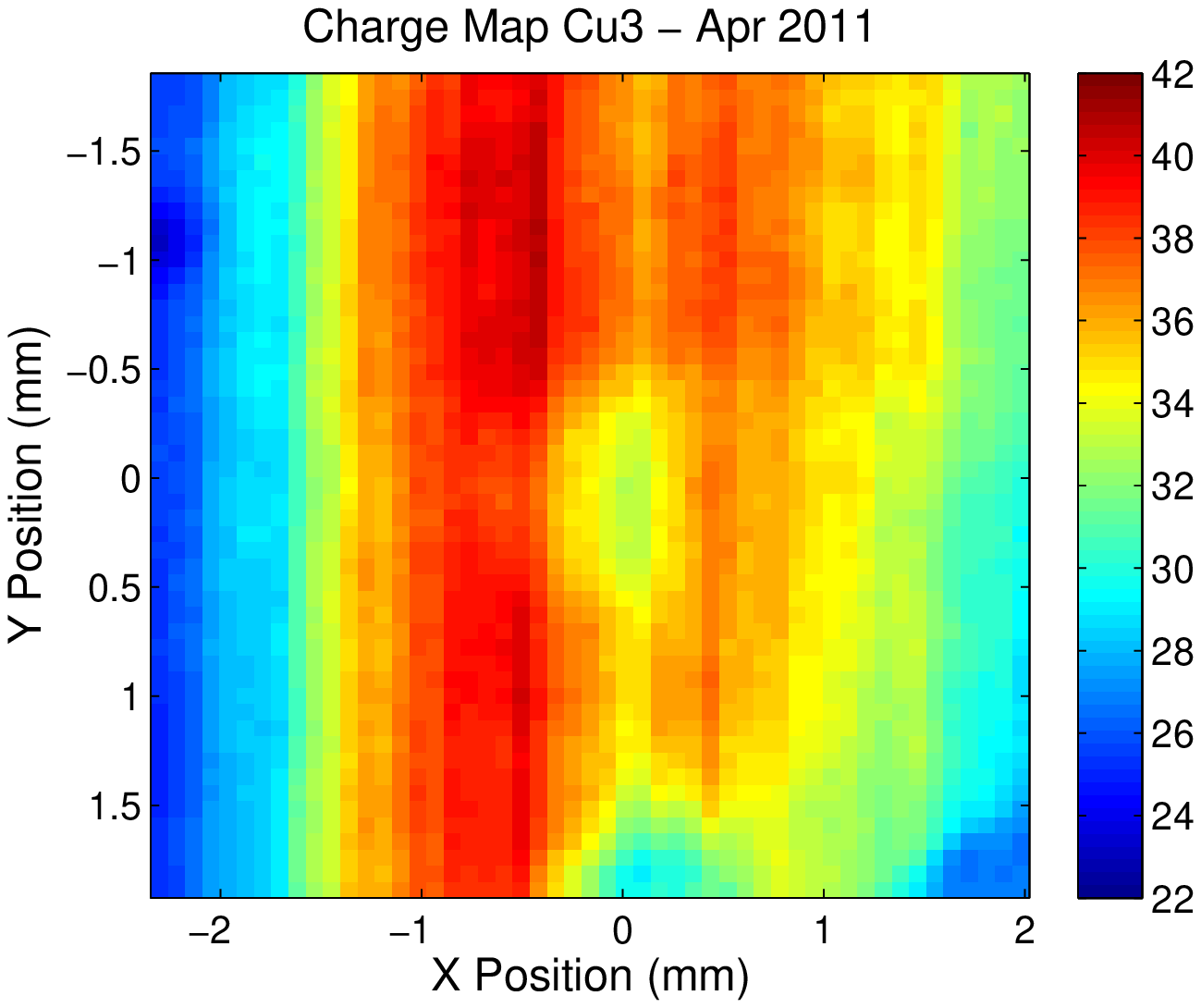}
\end{minipage}
     \caption{Charge map (pC) evolution of the Cu photocathode (Left plot: initial charge map)
     after three months of Injector operation (right plot).}
\label{fig250ChargeMap}
\end{figure}

\begin{figure}[htbp]
\centering
\includegraphics[width=0.8\textwidth, clip=]{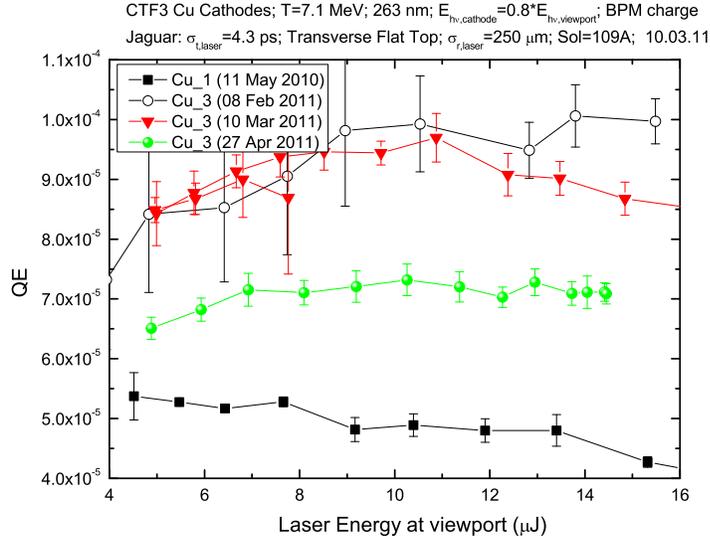}
     \caption{New Cu photocathode (Cu\_3) QE evolution after three months of Injector operation.
     Comparison with the QE of the first cathode installed Cu\_1 \cite{Ganter:private}.}
\label{figQECu1Cu3}
\end{figure}

The ageing of the cathode is characterized by the formation of a
QE hole, or a charge hole, as shown in Fig.\ref{fig250ChargeMap}.
Fig.\ref{figQECu1Cu3} shows the QE evolution in time of one Cu
cathode (Cu\_3) and is compared to the QE of another Cu
photocathode (Cu\_1) after a year of operation.

We were able to illuminate some Cu cathodes using a high repetition
rate UV laser producing a low fluence on the surface. What we
observed was an increase of QE with time. \newline At the PSI
injector, the UV lasers work at a low repetition rate (10~Hz); and
this until the end of the commissioning period. The two lasers used
at the injector Jaguar$^{\registered}$ or Pulsar$^{\registered}$ are
set to operate with a 10~ps long pulse, at wavelength of
$\lambda$=262~nm or 270~nm. The laser spot size is 1~mm on target.
We typically use 10-20 $\mu$J per pulse, hence a Fluence of 1.27 to
2.55~mJ/cm$^2$ and a power density of 127 to 255~MW/cm$^2$.

\begin{figure}[htbp]
\begin{minipage}[t]{.5\linewidth}
\centering
\includegraphics[width=0.9\textwidth,clip=]{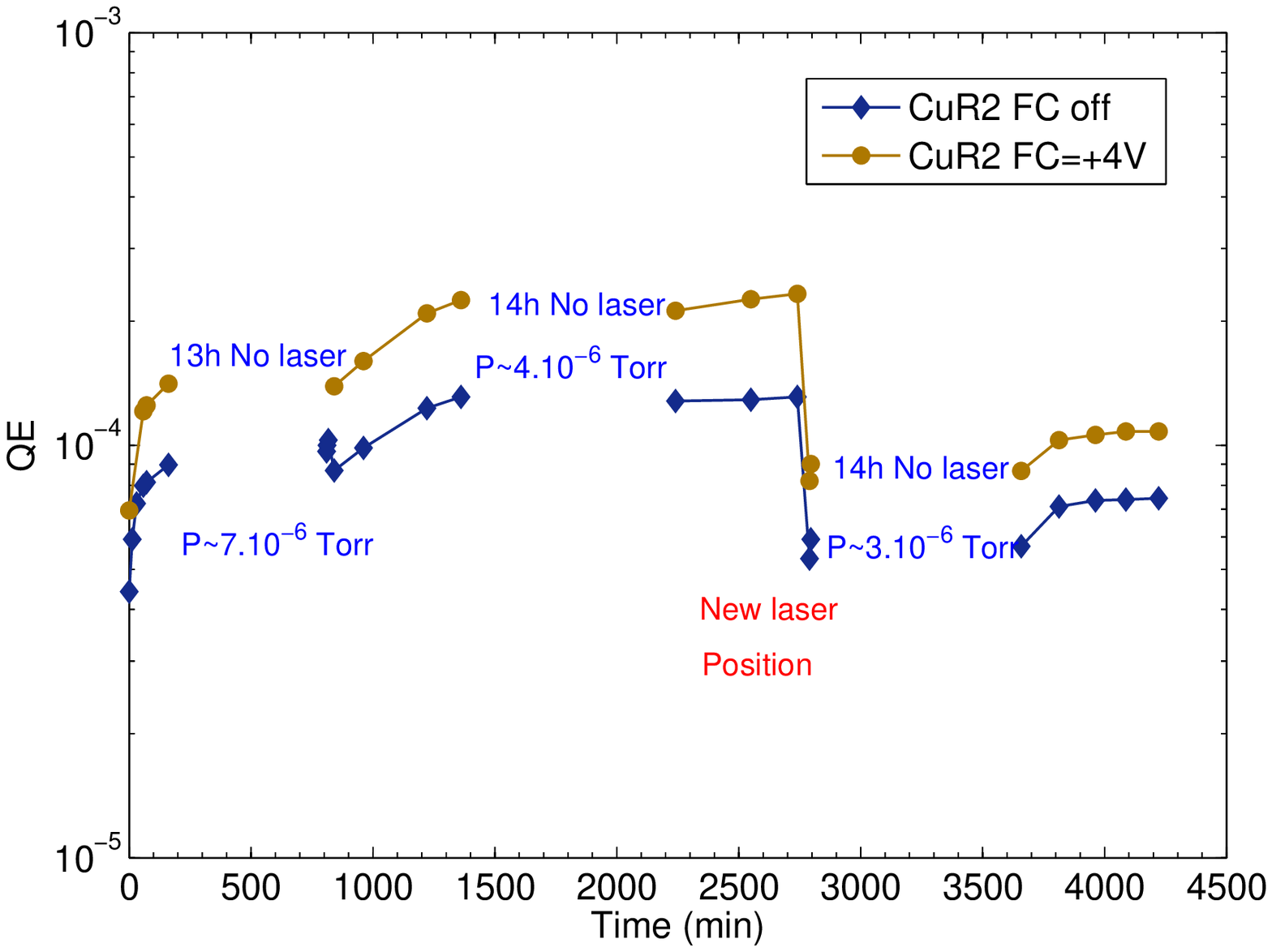}
\end{minipage}%
\begin{minipage}[t]{.5\linewidth}
\centering
\includegraphics[width=0.9\textwidth,clip=]{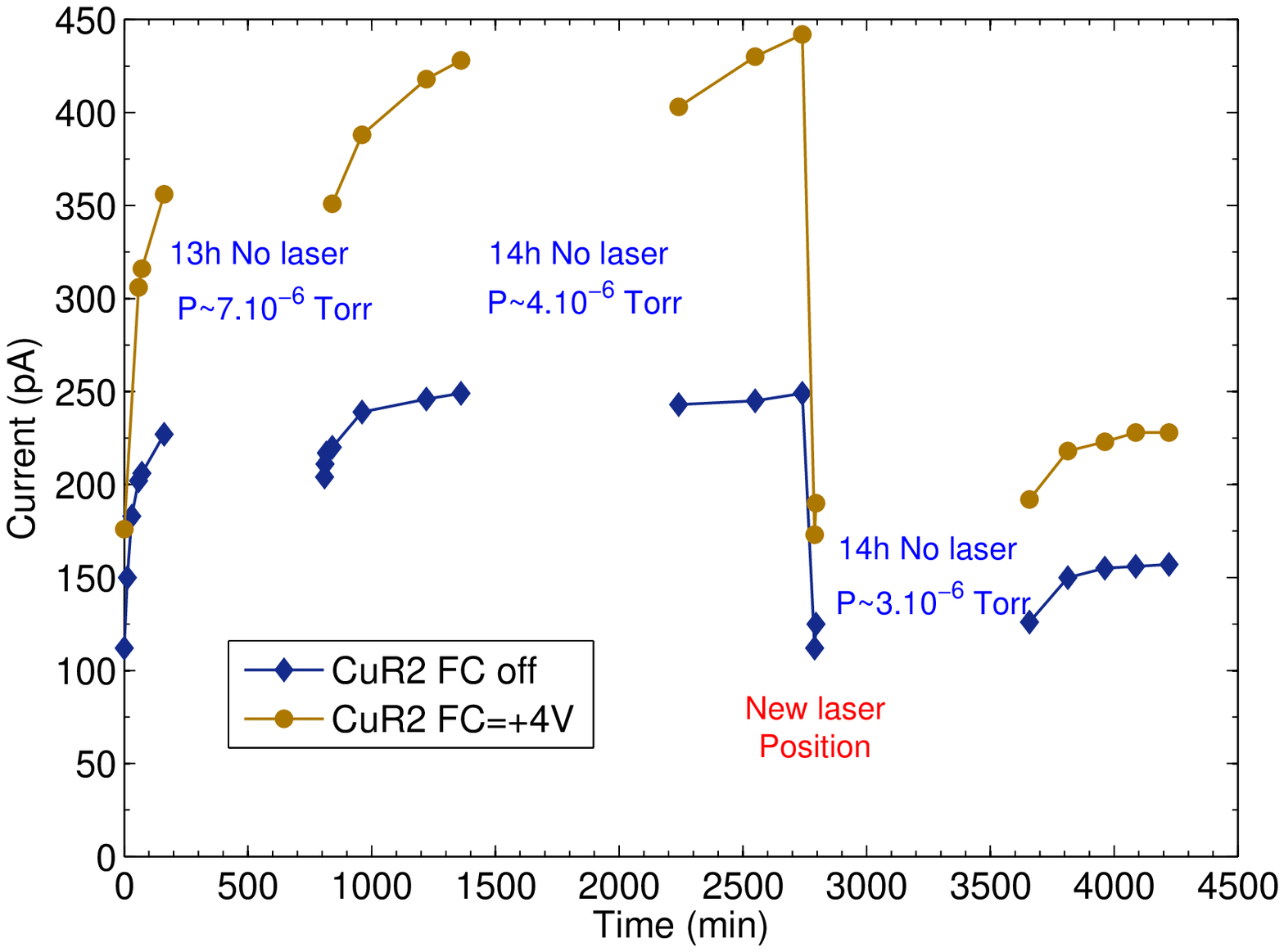}
\end{minipage}
     \caption{QE (left) and Current (right) measured from a Cu
     cathode (CuR2) using the Topas$^{\registered}$ laser
     with a 1~kHz repetition rate and with $\lambda$=261~nm}
\label{figTopasCuR2}
\end{figure}

In order to reproduce the QE hole in the small system chamber, we
have used the Topas$^{\registered}$ laser with the following
settings : 1000 Hz repetition rate, 1~ps long pulse,
$\lambda$=261~nm, a laser spot size between 1 to 1.2~mm on target.
The energy per pulse was 9 $\mu$J so a Fluence between 0.8 to
1.15~~mJ/cm$^2$ and a power density between 800 to 1150~MW/cm$^2$.
The QE and current measured during the laser exposure of a Cu insert
are shown in Fig.\ref{figTopasCuR2}. The already used Cu sample
(CuR2) was installed after an ethanol wipe. The sample was exposed
to the Duetto$^{\registered}$ laser 90~days before this experiment,
and was stored in air during that time. Any exposure to air cancels
any effects from a laser conditioning/cleaning.

The Topas$^{\registered}$ was not available during nights, hence the
gaps in the conditioning curve in Fig.\ref{figTopasCuR2}. The
injector operation is also usually stopped during nights. The
overall time of the laser on target is $\sim$20~hours. The fluence
in the small chamber is, in the worse case, 70\% of the injector gun
fluence. This amount for an equal time of 58~days of injector
operation. By that time, we should have seen a degradation of the
QE, and not an improvement as seen on the left plot of
Fig.\ref{figTopasCuR2}.
\newline We then moved the laser spot to another location on
the cathode and have further reduced the laser spot size to 0.7~mm.
The initial QE, t=2790~min, is similar to the first starting point
at t=0~min. The QE did increase but then seemed to level off after a
few hours of operation. The laser energy was constant to 10~$\mu$J
$\pm$ 1.5~$\mu$J. The sample CuR2 showed no surface damage upon
removal from the chamber.

We changed the sample by another Cu insert, which was also
previously used and stored in air. We irradiated this sample using
the Duetto$^{\registered}$ laser with its wavelength set to 266~nm,
with 200~kHz repetition rate. The spot size was set to 0.4~mm in
diameter, and the fluence used decrease from initially
255~$\mu$J/cm$^2$ to 65~$\mu$J/cm$^2$. At high fluence the sample
produced up to 7~nA of current with the FC set at +4~V and a couple
of nA with an off voltage on the FC. We probed a different location
on the insert with a lower laser fluence. The laser beam size was
sufficiently small to avoid any overlap on a previously irradiated
area. Again, we have seen no sign of QE degradation but on the
contrary a QE increase, even after 3 continuous days of irradiation
at 200~kHz (266~nm).

We can only hypothesize about the reasons why we could not degrade
the QE. Our RF photogun is baked to 120~C during almost a week.
Hence a different vacuum spectrum than an unbaked vacuum system. In
the baked case the vacuum spectrum is dominated by the hydrogen peak
(2~uma) and not by water. The cathode in the electron gun is
submitted to a high power (20~MW) and high gradient (100~MV/m) of RF
field alternating at 3~Ghz. This RF field produces dark current and
ions. Both can strike back the cathode and degrade the QE. However,
Fig.\ref{figQECu1Cu3} (right plot) still shows a mostly homogeneous
QE over the whole cathode.
\newline The vacuum environment of our system is at first dominated by
water ($>$95\%). It is then possible that the intense laser light
cracks the water molecules, producing very reactive radicals. These
radicals would continuously restore the QE, as does an Ozone
cleaning \cite{Lecce:2011}. The pressure is usually in the
10$^{-6}$~Torr range. At this pressure a monolayer is formed every
second. Laser heat can also activate surface molecule diffusion, and
maybe the re-arrangement at this pressure of the oxide layer is
beneficial, although the mechanisms are unclear. However, one should
note that the vacuum spectrum of an unbaked system can be similar,
qualitatively, to a baked system; if one allows the vacuum chamber
to be pumped long enough.

\section{Conclusion}

It seems a good idea to measure the work function of a technical
metal using an infrared laser coupled with an optical parametric
amplifier. However, as we have seen in Fig.\ref{figQECuR1R2} the QE
(or extracted electron current) decreases but do not drop sharply
for wavelength longer than the work function. This would be what one
would expect after experimenting with a powerful UV lamp coupled to
a monochromator. Double-photon absorption leading to photoemission
was put in evidence for a UV laser with a long emission pulse
(10~ps) on Al, AlLi, Cu, and Mg; ($\lambda_{Laser} > \lambda_{WF}$).
We were not able to determine the cause of the photoemission when
using ps or shorter laser pulse length.

We have observed that a UV (266~nm and 355~nm) high repetition rate
laser (1~kHz or 200~kHz) shined on various metallic surfaces located
in an unbaked UHV vacuum environment is beneficial to the QE, as it
increases it. These results are in agreement with the results
obtained using aggressive laser cleaning procedure applied on
photocathodes for RF photoguns. We observe, contrary to laser
cleaning techniques, that the mirror-like polished surfaces stay
pristine after our extensive but not aggressive laser exposure.

We were not able to reproduce the QE hole which is commonly seen on
RF photoguns' photocathodes after many hours of operation, in spite
of the use of laser parameters similar to parameters used for
electron production at the PSI photogun. We attribute this fact to
probably the most important parameter, the absence of the RF field.
Secondly to the difference of vacuum composition between a baked RF
photogun, which mainly contains hydrogen, and our unbaked system
which predominantly contains water. We hypothesize that the water
molecules are cracked by the laser beam on the photocathode and that
the radicals produced enhance the QE instead of being detrimental to
it.

\section{Acknowledgments}

S.~Ivkovic for polishing the numerous test samples, sometimes on
short notice. The SLS vacuum group and pulse magnet group for
lending some equipments and providing knowhow. The FEMTO group for
allowing us to use their lasers during the SLS shutdown. M.~Divall
for valuable discussion and F.~Celli for proofreading this text.

%
%



-----------------------------------------------
%
%
\clearpage
\listoftables
\newpage
\listoffigures

\end{document}